\documentclass[a4paper,11pt]{article}
\usepackage{pos}
\usepackage{hyperref}
\usepackage{siunitx}

\usepackage{xspace}
\newcommand{\cor}[1]{CORSIKA~{#1}\xspace}

\title{CORSIKA 8: A modern and universal framework for particle cascade simulations}

\author*[a]{Marvin Gottowik}
\onbehalf{for the CORSIKA 8 collaboration\footnote[2]{email: \href{mailto:corsika8@kit.edu}{\texttt{corsika8@kit.edu}}, full author list: \url{https://gitlab.iap.kit.edu/AirShowerPhysics/corsika/-/wikis/UHECR24-CORSIKA-8-author-list}}}

\affiliation[a]{Karlsruher Institut für Technologie, Institut für Astroteilchenphysik, Karlsruhe, Germany}

\emailAdd{marvin.gottowik@kit.edu}

\abstract{\cor8 represents a significant update in the simulation of particle showers, building on the well-established foundation of \cor7. It has been entirely rewritten as a modular and modern C++ framework, addressing the limitations of its predecessor to provide a flexible platform designed to satisfy current and novel use cases. This allows for application beyond pure air-shower scenarios such as cross-media particle cascades and an advanced calculation of the radio emission. A first official ``physics-complete'' version has already been released that supports the treatment of hadronic interactions with Sibyll 2.3d, QGSJet-II.04, and EPOS-LHC and the treatment of the electromagnetic cascade with PROPOSAL 7.6.2. In this presentation, we will discuss the design principles, current functionality, and validation efforts of \cor8, emphasizing its potential applications for future experiments.}

\FullConference{
7th International Symposium on Ultra High Energy Cosmic Rays (UHECR2024)\\
17-21 November 2024\\
Malargüe, Mendoza, Argentina}

\begin{document}
\maketitle

\section{Introduction}

Monte Carlo simulations are essential in the field of astroparticle physics for the interpretation of measured data. CORSIKA (latest version 7.7550) has been widely used for more than 20 years to simulate air showers initiated by ultra-high-energy cosmic rays in the Earth's atmosphere. While \cor7 has served the community well, its design is limited to atmospheric showers, making it unsuitable for simulations in other media such as water or ice, which are crucial for modern experiments. To meet these evolving needs, the development of \cor8 was launched in 2018 as a complete rewrite within a flexible and modular C++ framework. \cor8 not only extends the simulation capabilities beyond air showers to include particle cascades in arbitrary media but also enables the simulation of cross-media showers, a key feature for a broader range of current and future experiments. The \cor8 code is considered ``physics complete'' and a first expert version of it was released at the end of 2024 \cite{c8_release}. We are now focused on making \cor8 more accessible by developing a containerized version that users can easily run out of the box. We are also working on enhancing aspects such as performance, simulation steering, and documentation to improve the overall user experience.

The general design of \cor8 is described in reference \cite{ng_corsika}. In this proceeding, we present the latest developments in \cor8, focusing on the validation of air showers against \cor7, the inclusion of radio emission calculations, and new functionality such as cross-media shower simulations. Additionally, we introduce new tools, a Python library for reading \cor8 output and a graphical interface for visualizing simulations, designed to enhance usability for the community.

\section{Air shower validation}

To ensure the reliability of \cor8, we validate its simulation results against those of \cor7. As a first step, we compare electromagnetic showers initiated by a 100 TeV primary electron, cf.\ reference \cite{c8_EM} for details. Examining the longitudinal and lateral shower profiles as well as the energy distribution, the differences typically remain within \SI{10}{\percent}. This level of agreement is remarkable given that different codes are used to describe the electromagnetic cascade: PROPOSAL \cite{koehne2013proposal, dunsch_2020_1484180, dunsch_2018_proposal_improvements} in \cor8 versus a customized version of EGS4 in \cor7.

We now extend our comparison to hadronic showers. \cor8 employs a modular framework that integrates state-of-the-art hadronic interaction models. At high energies, it currently supports EPOS-LHC~\cite{eposlhc}, Sibyll 2.3d~\cite{sibyll}, and QGSJet-II.04~\cite{qgsjet}, as well as a preliminary implementation of Pythia 8~\cite{pythia, c8_pythia, c8_pythia2}.  Ongoing efforts also focus on integrating updated models such as EPOS LHC-R~\cite{eposlhcr} and QGSJet-III~\cite{qgsjet3}. Low-energy interactions are handled using FLUKA~\cite{FLUKA3}.

\begin{figure}
\centering
\includegraphics[width=0.5\linewidth]{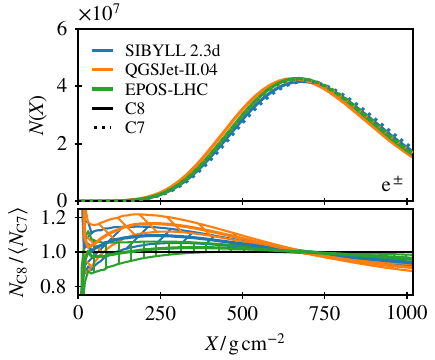}%
\includegraphics[width=0.5\linewidth]{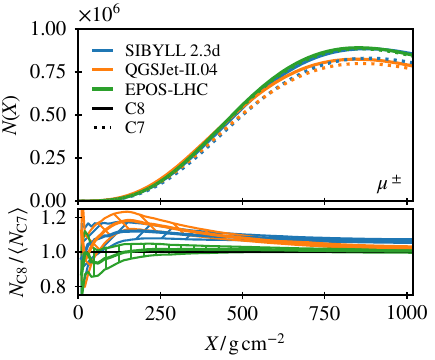}

\includegraphics[width=\linewidth]{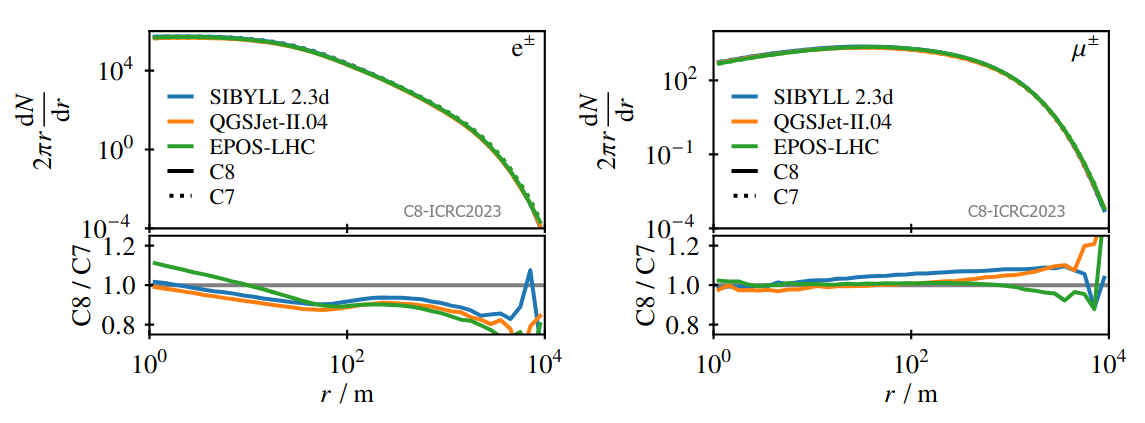}
\caption{Average longitudinal profile (top) and lateral distributions (bottom) of electrons plus positrons (left) and muons (right) at ground for 300 vertical proton-induced air showers with a primary energy of \SI{e17}{eV}. The hatched area shows the standard deviation of the mean. Figures from \cite{ICRC-C8}.}
\label{fig:lateral_profile}
\end{figure}

In Fig.\ \ref{fig:lateral_profile}, we showcase the average longitudinal profile and lateral distribution of electrons plus positrons and muons for 300 vertical proton-induced air showers with an energy of \SI{e17}{eV}, cf.\ reference \cite{ICRC-C8} for details on the exact simulation setup. We observe a typical agreement for various particle species on the $\sim$\SI{10}{\percent} level. The observed differences seem to be dependent on the high-energy interaction model. However, reaching agreement between completely independent codes at this level is a notable accomplishment and further investigations will be conducted to better understand the origin of these discrepancies.

\section{Radio emission calculations}
Radio-emission calculation has been a key driver in the development of \cor8, primarily due to the radio-detection community's need for flexibility in handling more complex scenarios, such as radio emission from air showers crossing into ice for in-ice detection experiments, cf.\ following section. To address this, \cor8 decouples the emission calculation from the signal propagation. This design enables easy integration of specific use cases required by the community. \cor8 can perform concurrent simulation of the radio emission for the identical particle cascade with two established formalisms: the endpoint formalism as originally implemented in the \cor7 extension CoREAS~\cite{CoREAS} and the ZHS formalism available in ZHAireS ~\cite{ZHAiRES}.

A comparison of radio-emission predictions between \cor8 using both formalism and the established CoREAS and ZHAireS codes showed very good qualitative agreement, particularly in the symmetry and polarization of the energy fluence footprints. For high-precision simulations, the difference in the radiation energy, the energy emitted by charged particles in the form of electromagnetic waves, reaches less than \SI{10}{\percent} in the \SIrange{30}{80}{MHz} band currently used by many experiments and becomes negligible in the \SIrange{50}{350}{MHz} band. A detailed explanation of the implementation of the radio process in \cor8 and an in-depth analysis of the predicted radio emission is presented in reference \cite{radio_paper}.

\section{Cross-media showers}
One of the most significant advancements in \cor8 is its ability to simulate cross-media showers, a functionality that previously required the use of multiple tools. In the past, simulating particle cascades that transition between different media, such as air into ice or water, was a complex task involving the combination of \cor7 for atmospheric showers and GEANT for simulating interactions in dense media. This approach was cumbersome and limited by the need to manually integrate results from different codes, each with its own set of assumptions and interfaces. \cor8 now brings everything into a single, unified framework, making it possible to simulate showers that span multiple media seamlessly. The modular structure of \cor8 allows users to define complex environments where different materials can interact with incoming high-energy particles, providing a more accurate and consistent picture of the particle cascade process. This capability is crucial for current and future experiments, such as  ultra-high energy neutrino detection.

\begin{figure}
\centering
\includegraphics[width=0.5\linewidth]{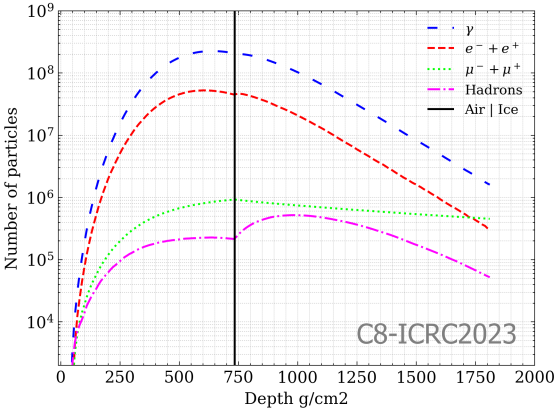}%
\includegraphics[width=0.5\linewidth]{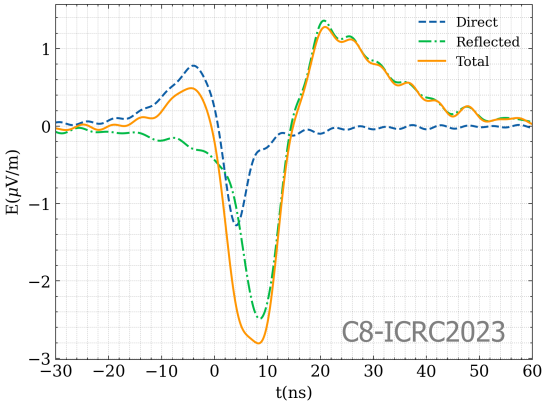}
\caption{Comparison of a \SI{100}{PeV} proton shower propagating through the atmosphere and intersecting an ice core at \SI{2.4}{km}. On the left, the longitudinal profile simulated with \cor8 is shown, and on the right, the radio emission from the in-ice portion of the shower, considering direct and reflected optical paths. Figures from \cite{icrc_cross_media}.}
\label{fig:cross_media}
\end{figure}

As an example, we compare a \SI{100}{PeV} proton shower propagating through the atmosphere and intersecting an ice core at an altitude of \SI{2.4}{km}, cf.\ reference \cite{icrc_cross_media} for details. The agreement between \cor8 (shown in Fig.\ \ref{fig:cross_media} (left)) with previous work using \cor7 and GEANT4 is generally good, with the key difference being that \cor8 accounts for the hadronic interactions in the ice. This rehadronization process in \cor8 leads to an increase in hadrons and a small bump in electron and photon profiles as neutral pions decay into photons after the interface.

We now explore the simulation of radio emission from the in-ice part of the shower. The radio emission can travel via two possible optical paths: directly through the ice or reflected at the ice surface. The results are shown in Fig.\ \ref{fig:cross_media} (right). This application is made possible by the flexibility of \cor8's radio module, which allows for the treatment of complex media transitions. We note that this is a simplified treatment for the purpose of testing the radio interface. At this stage, the simulation only accounts for the basic signal paths, without incorporating more complex effects like phase shifts that might occur at the reflections.

Additional applications of simulating in-ice radio emissions from air showers using \cor8 are discussed in \cite{alan}.

\section{Python library and the C8 viewer}
The \cor8 output is organized into separate directories for each observing process. At the top level, human-readable YAML files store configuration details and summary information, both globally and for each observing process. The global YAML file includes details such as the \cor8 version, exact steering parameters, random seed, and runtime. Observer-specific YAML files provide additional metadata, such as the units used or, in the case of radio simulations, the locations where radio emission was calculated. The primary data from each observing process is stored in a Parquet file. Although these files can be read manually, we recommend using the dedicated Python library included with \cor8. This library serves as an abstraction layer, ensuring that analysis scripts remain functional and independent of specific output formatting, even if the structure evolves in future versions. In addition, example scripts are provided that demonstrate how to use the library for data analysis to help users get started.

\begin{figure}
\centering
\includegraphics[width=\linewidth]{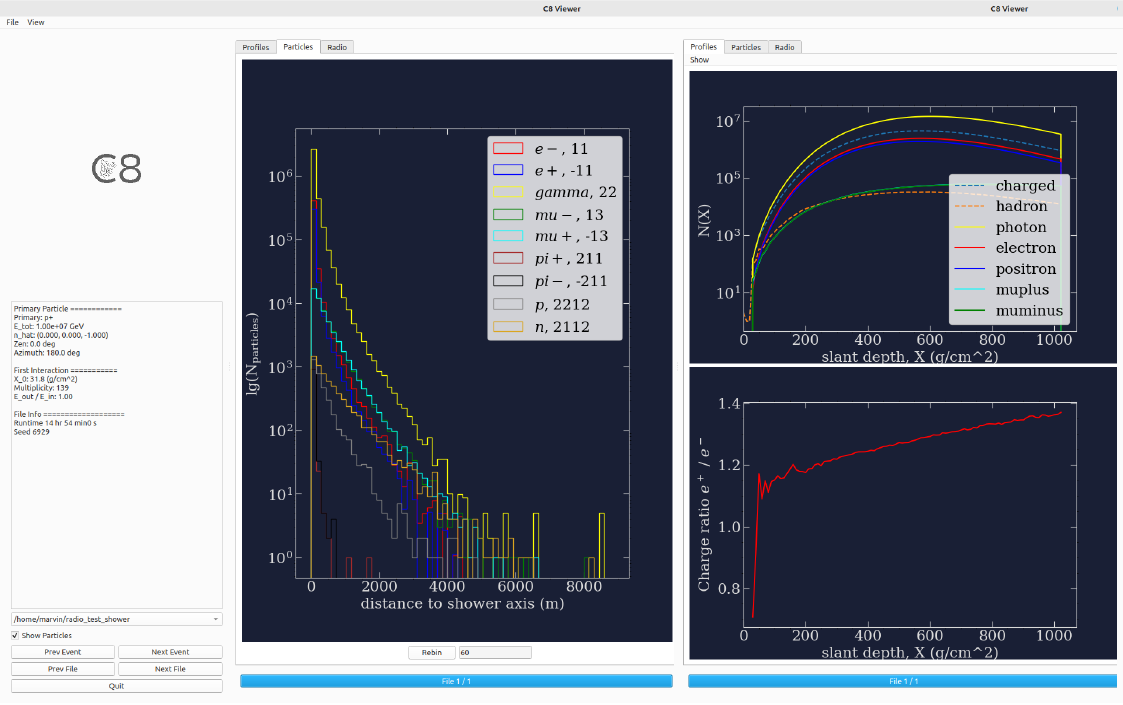}
\includegraphics[width=\linewidth]{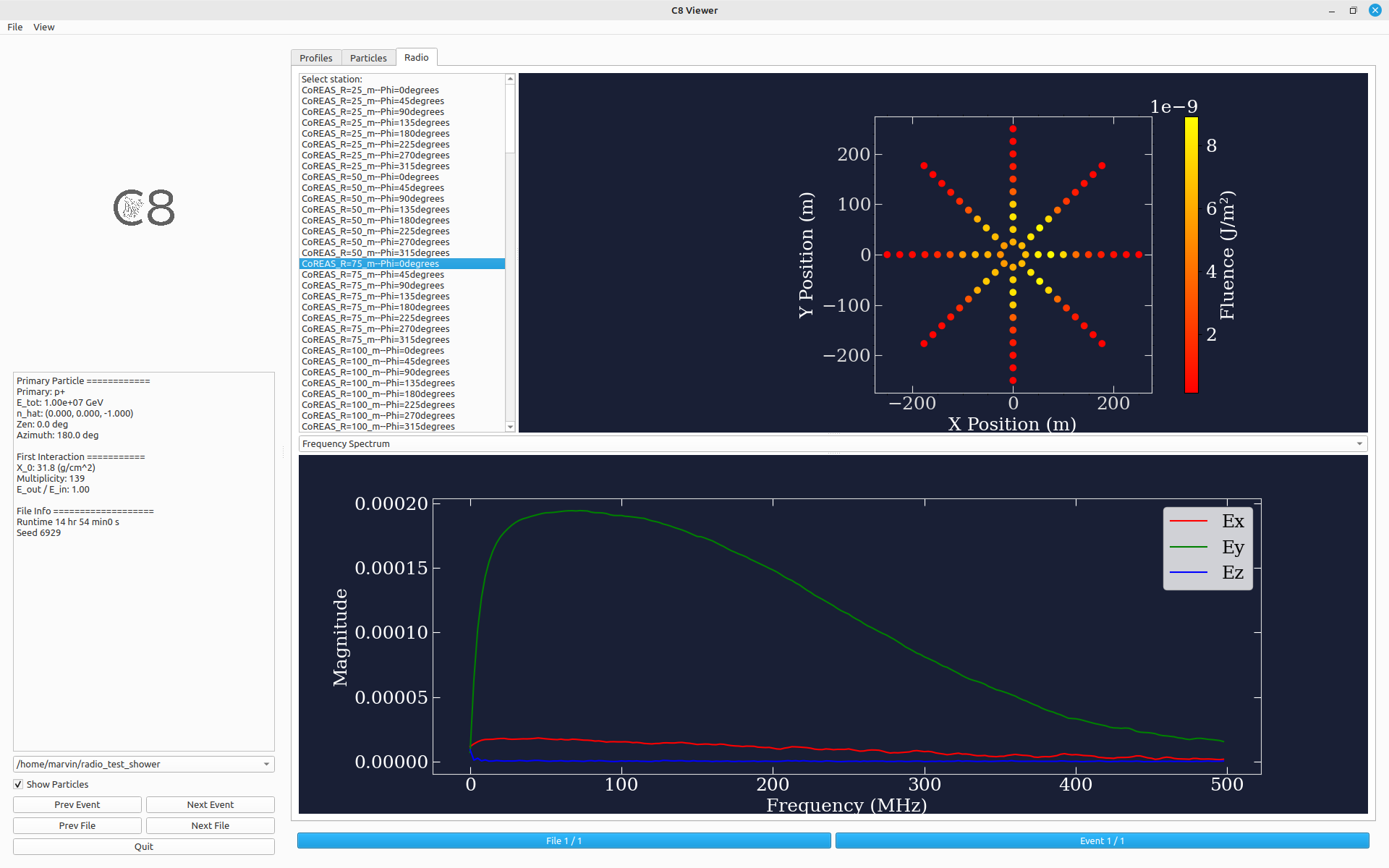}
\caption{Screenshots of the ``C8 Viewer''. On the left hand side general information on the simulated air shower are displayed. The top plot contains a combination of the ``Profile'' and ``Particles'' tab visualizing the lateral distribution and longitudinal profile of different particle types as well as the ratio of electrons and positrons. The bottom figure contains the ``Radio'' tab, showing the time trace or frequency spectrum of simulated pulses and a view of the distribution of the radio emission in the shower plane.}
\label{fig:c8_viewer}
\end{figure}

To further enhance user experience, \cor8 will be accompanied by the ``C8 Viewer'', a graphical tool designed for intuitive and interactive visualization of simulated events, as shown in Fig.\ \ref{fig:c8_viewer}. This GUI-based application enables users to explore various aspects of particle cascades in detail. General information about the primary particle, the first interaction, and simulation runtime is displayed in a dedicated text panel on the left. On the right, multiple tabs provide information on key aspects of the shower development: the longitudinal profile, lateral particle distributions, and radio emission characteristics.

The ``Profiles'' tab presents the number of particles as a function of amospheric depth for different particle types, the energy deposited along the shower axis, and the electron-to-positron ratio. The ``Particles'' tab visualizes the lateral distribution of various particle species at the ground, including the lateral footprint of the shower and the distribution of particle weights from the thinning. The ``Radio'' tab provides a list of all simulated pulse locations, displaying the time-domain signal and its frequency spectrum for individual stations. Additionally, it computes and visualizes the distribution of energy fluence, the energy carried by radio waves per unit area, for all observer positions. For specialized star-shape simulations in the shower plane, the energy fluence map allows interpolation of the full radio footprint, revealing the expected emission features. These include the interplay between geomagnetic and charge-excess contributions and the characteristic Cherenkov-like compression of the radio signal into a ring around the shower axis \cite{th-radio}.

\section{Conclusion}

Over the past years, \cor8 has made significant progress, reaching a stage where it is considered ``physics-complete.'' This means that the fundamental physical processes governing particle shower development and the calculation of the radio emission are fully implemented. Extensive validation against \cor7 has demonstrated agreement at the $\sim$\SI{10}{\percent} level, with remaining differences under active investigation. An expert version of it was released at the end of 2024 to obtain broader feedback from the community \cite{c8_release}. For the simulation of particle showers developing in dense media or crossing from air into dense media, we consider \cor8 the code of choice already now. However, \cor7 remains the recommended choice for standard air-shower applications at this time.


\begin{thebibliography}{99}
\sloppy

\def\vyp#1#2#3{\textbf{#1} (#2) #3} 

\bibitem{c8_release}
\emph{CORSIKA 8 - Release v1.0-beta1}, \url{https://gitlab.iap.kit.edu/AirShowerPhysics/corsika/-/tags/corsika8-v1.0-beta1}.

\bibitem{ng_corsika}
R. Engel, D. Heck, T. Huege et al.,  Comput. Softw. Big Sci. \vyp{3}{2019}{2}.

\bibitem{c8_EM}
A. Sandrock, J-M. Alameddine and F. Riehn for the CORSIKA 8 collaboration, PoS(ICRC2023)393.

\bibitem{koehne2013proposal}
J.-H. Koehne, K. Frantzen, M. Schmitz et al., Comput. Phys. Commun. \vyp{184}{2013}{2070}.

\bibitem{dunsch_2020_1484180}
J.-M. Alameddine, M. Dunsch, L. Bollmann et al., \emph{tudo-astroparticlephysics/PROPOSAL: Zenodo} (2020).

\bibitem{dunsch_2018_proposal_improvements}
M. Dunsch, J. Soedingrekso, A. Sandrock et al., Comput. Phys. Commun. \vyp{242}{2019}{132}.

\bibitem{eposlhc}
T. Pierog, I. Karpenko, J.M. Katzy, E. Yatsenko and K. Werner, Phys. Rev. C \vyp{92}{2015}{034906}.

\bibitem{sibyll}
F. Riehn, R. Engel, A. Fedynitch, T.K. Gaisser and T. Stanev, Phys. Rev. D \vyp{102}{2020}{063002}.

\bibitem{qgsjet}
S. Ostapchenko, Phys. Rev. D \vyp{83}{2011}{014018}.

\bibitem{pythia}
T. Sjöstrand and M. Utheim, Eur. Phys. J. C \vyp{82}{2022}{21}.

\bibitem{c8_pythia}
C. Gaudu, M. Reininghaus, and F. Riehn for the CORSIKA 8 collaboration, arXiv:2412.15094 (2024).

\bibitem{c8_pythia2}
C. Gaudu for the CORSIKA 8 collaboration, PoS(UHECR2024)089.

\bibitem{eposlhcr}
T. Pierog and K. Werner, PoS(ICRC2023)230.

\bibitem{qgsjet3}
S. Ostapchenko, arXiv:2410.23465 (2024).



\bibitem{FLUKA3}
F. Ballarini, et al., EPJ Nucl. Sci. Technol. \vyp{10}{2024}{16}.

\bibitem{ICRC-C8}
T. Huege and M. Reininghaus for the CORSIKA 8 collaboration, PoS(ICRC2023)310.

\bibitem{CoREAS}
T. Huege, M. Ludwig and C.W. James, AIP Conf. Proc. 1535 (2013) 128.

\bibitem{ZHAiRES}
J. Alvarez-Muñiz, W.R. Carvalho Jr. and E. Zas, Astropart. Phys. 35 (2012) 325.

\bibitem{radio_paper}
J.M. Alameddine, et al., Astropart. Phys. \vyp{166}{2025}{103072}.

\bibitem{icrc_cross_media}
J. Ammerman-Yebra for the CORSIKA 8 collaboration, PoS(ICRC2023)442.

\bibitem{alan}
A. Coleman, C. Glaser, R. Rice-Smith, et al., arXiv:2410.08615 (2024).

\bibitem{th-radio}
T. Huege, Physics Reports, \vyp{620}{2016}{1}.

\end{thebibliography}
\end{document}